# Sub-10 nm colloidal lithography for integrated spin-photo-electronic devices

*A. Iovan, M. Fischer, R. Lo Conte, and V. Korenivski*

*Nanostructure Physics, Royal Institute of Technology, 10691 Stockholm, Sweden*

**Colloidal lithography [1] is how patterns are reproduced in a variety of natural systems and is used more and more as an efficient fabrication tool in bio-, opto-, and nano-technology. Nanoparticles in the colloid are made to form a mask on a given material surface, which can then be transferred via etching into nano-structures of various sizes, shapes, and patterns [2,3]. Such nanostructures can be used in biology for detecting proteins [4] and DNA [5,6], for producing artificial crystals in photonics [7,8] and GHz oscillators in spin-electronics [9-14]. Scaling of colloidal patterning down to 10-nm and below, dimensions comparable or smaller than the main relaxation lengths in the relevant materials, including metals, is expected to enable a variety of new ballistic transport and photonic devices, such as spin-flip THz lasers [15]. In this work we extend the practice of colloidal lithography to producing large-area, near-ballistic-injection, sub-10 nm point-contact arrays and demonstrate their integration in to spin-photo-electronic devices.**

Patterning of materials at sub-10 nm dimensions is at the forefront of nanotechnology and employs techniques of various complexity, efficiency, areal scale, and cost. Electron beam and focused ion beam techniques are typically limited to feature sizes of tens of nanometers, if the features are to be well defined, and are rather inefficient for large-area nano-patterning since both methods employ series point-by-point pattern transfer. Two promising techniques of nano-imprint lithography [16,17] and extreme ultra-violet interference lithography do indeed open for exploration sub-10 nm nanostructures [18]. The instrumentation required, however, can in many cases be of great complexity and cost. Recently, membranes of nano-porous anodic aluminum oxide [19] were shown to scale to sub-10 nm dimensions and potentially compete with the above advanced lithographic techniques at this scale. Another potential alternative for sub-10 nm patterning is colloidal lithography, which is very attractive at larger dimensions due to its ease of use and low cost. Colloid-based patterning is known to be capable of producing individual sub-10 nm objects. However, ordered large-area nano-arrays fully integrated into photonic or electronic devices have not been demonstrated using colloidal lithography. In this work we use a self-assembled monolayer of polystyrene nanoparticles, reduced in size by a novel etching process, which enables sub-10 nm feature sizes with large-area coverage in a well-defined hexagonal lattice and full integration for electrical circuit biasing and read out. We demonstrate the new fabrication technique using spin-torque and spin-flip photoemission material combinations, considered promising for GHz oscillators and THz lasers.

## Self-assembled monolayer of nanoparticles

The most widely used colloidal lithography medium is polystyrene nanoparticles in aqueous solution. Such colloidal solutions are commercially available with a variety of concentrations and particle sizes [20]. We have used a range of diameters (down to 40 nm) and found the most consistent results in terms of self-assembly for 200 nm diameter and 2% particle concentration. There exist different methods of forming a monolayer of colloidal particles on a surface [21]. We found spinning the polystyrene colloidal water solution to yield good results. In calibrating the speed and duration of the spinning we aimed at forming the largest-area continuous monolayer possible. Thus, spinning in three stages, 500 rpm for 10s, 1000 rpm for 30s, and 2000 rpm for 10 s, facilitates self-assembly and yields continuous nanoparticle monolayers of hundreds of micrometers in area. This was sufficient for our purposes to demonstrate a wide range of integrated device sizes.

The first step of the spinning sequence allowed the particles to gently segregate onto the substrate and, to a large extent already here, form a hexagonal pattern. The second step of spinning at a higher rpm-speed prevented formation of additional layers of vertically stacked particles. The last step was used to remove the remaining solution, predominantly in the corners of the sample.

The spinning process was developed using $SiO_2$ substrates, about 2x2 $cm^2$, covered with a 10 nm thick Au layer. Prior to spinning the substrate was etched in plasma oxygen for 2 minutes in order to make the surface hydrophilic [22]. As an alternative route, we found that a good quality monolayer, with a well-defined long-range order, can be obtained if a small amount of Triton X surfactant [23] is added into the colloidal solution. However, the subsequent extensive tests showed that the use of Triton

X is problematic for later processing during the device integration steps. Traces of Triton remaining on the surface produce residue that prevents reliable integration. We therefore selected the route of forming large-area self-assembled monolayer arrays of nanoparticles by making the substrate hydrophilic with the help of RIE plasma oxygen.

Figures 1a and 1b show the tapping mode atomic force microscopy (AFM) images of a typical monolayer, with the particle diameter (and the inter-particle distance) of 200 nm. The monolayer film is of good quality, with only minor defects on the large scale (Fig. 1a). On the small scale (Fig. 1b) the lattice clearly is hexagonal. The scanning electron microscopy (SEM) image of the sample in Fig. 1c shows that the nanoparticle array has a nearly perfect close-packed hexagonal lattice. The dispersion in the particle size at 200 nm diameter is small (approximately 1%), which favors the translation of the hexagonal pattern over large areas, hundreds of microns in the case of our optimized self-assembly process.

## Downscaling to sub-10 nm

Our process of down scaling the particles of the polystyrene monolayer to the 10-nm range consists of 4 main steps, illustrated in Fig. 2. Once the monolayer is formed (Fig. 2a), reactive oxygen plasma is used to reduce the size of the particles (Fig. 2b). When the desired particle diameter is reached, a reinforcing layer of Aluminum is deposited, as shown in Fig. 2c, which in the later process stages acts as a hard mask. A lift-off of the particles by etching completes the fabrication of the nano-mask, as illustrated in Fig. 2d. The very sensitive process step of the downscaling the particles is reactive plasma etching, which must be done in a very clean chamber [23] in order to have a uniform reduction in the particle size across the large-area array. The final particle diameter is found to be a smooth function of the etching time, so the feature size of the nano-mask can be controlled rather precisely. The typical etching power used is relatively low (50 W) to avoid potential disruptive etching at higher power. The key process detail, that we found to be critical for achieving sub-10 nm resolution, is a superposed inducted coupled plasma (ICP) of relatively high power (250 W), which increases the ionization in the chamber, translating into a more isotropic reduction in the particle size. As a result, the polystyrene particles remain nearly spherical during the process, even as their size is reduced by more than an order of magnitude. The Oxygen pressure was found to be optimum at 40 mT.

A typical monolayer after the etching procedure is shown in Fig. 3, imaged by AFM for etching quality, surface topography, and particle size. The height of the particles is measured accurately, but not the diameter since the convolution of a small particle and the tip produces a widening distortion. Keeping all the process parameters constant and varying only the plasma-ICP etching time, we reproduce the general result of previous studies [2] of a gradual reduction in the particle size. Figures 3a,b,c, show the particle monolayer after etching for 165, 180, and 195 seconds, respectively. The key advance made in this work, compared to the results on colloidal lithography reported to date, is that our modified process scales to sub-10 nm dimensions. For example, for the etching time of 2 min 45 s (Fig. 3a) the particle diameter is reduced to 10-15 nm (measured by AFM with particle-tip de-convolution). For 3 min etching time the average diameter is below 10 nm (Fig. 3b). Etching for 3 min 15 s reduces the average size further but induces some perturbations. Already at 3 min 30 s etching time the monolayer is significantly over-etched and the pattern is partly removed. The true size of the particles can be estimated using AFM traces, such as those shown in Fig. 3d. The size of the particles here is smaller than the actual curvature of the AFM tip, obtained from scanning calibration samples to be approximately 50 nm. Therefore, a de-convolution procedure was used to obtain the particle sizes stated above, which compare well with those obtained using SEM (see below). The AFM height is a more direct measurement and shows approximately 30 nm for 2 min 45 s and 15 nm for 3 min etching time. Figure 3 thus illustrates the fine control of the particle size at ~10 nm by varying the plasma etching time.

The technological viability of the obtained polystyrene nano-particle array depends on the ability to transfer it into a reliable mask to be used in subsequent nano-device integration. The most straightforward approach, used widely in the literature for larger particle-size patterns, would be to directly etch the underlying substrate (e.g., $SiO_2$ or Au) using the particle-array as the mask. Our extensive tests showed, however, that the particle-mask itself is significantly modified during this process, which makes the pattern transfer at the desired 10-nm-diameter range essentially impossible. We therefore developed an additional lift-off process step to reinforce the mask. It includes a deposition by e-beam evaporation of a 15 nm thick Al metal layer onto the etched particle array, with a subsequent lift-off step to remove the polystyrene particles, and yields the hard mask illustrated in Fig. 2d. Chemically removing the particles using acetone and mechanical polishing did not work. Another technique tested with only partial success was heating the sample just bellow the melting point of polystyrene, where the dilatation coefficient makes the particles expand in volume significantly and thereby break-open holes in the Al film. This is a promising technique, however, we found it difficult to control the size and shape of the resulting ~10-nm holes opened by polystyrene exploding through Al. A successful and stable lift off process at these small length-scales was found to be reactive-ion-etching (RIE) with Oxygen, in which the polystyrene particles are first etched predominantly from the sides, where the Al film is much thinner due to the shadowing effect of the e-beam coating of the polystyrene spheres. During this RIE etching step the Al film surface oxidizes and forms a hard mask for subsequent ion milling. After a 5 min Ar-plasma etch to remove surface residue, 7 min long ion milling etches through the 10 nm thick Au layer and slightly into the $SiO_2$ substrate, thus transferring the hexagonal pattern of sub-10 nm polystyrene particles into sub-10 nm pattern of holes in Au on $SiO_2$. The Al-oxide layer acts as the hard mask in this process.

Figure 4 shows SEM images of a typical nano-hole array mask. Long-range order is maintained over the micrometer range, as shown in Figs. 4a,b. The process was repeated many times and showed good reproducibility. The average inter-dot distance is 200 nm, corresponding to the original particle diameter (Figs. 4c,d) and the average hole size reaches down to the sub-10 nm range. The SEM data is well calibrated and confirms the de-convoluted AFM data discussed above as regards the morphology of the array and the size of the particles, throughout the process.

## Device integration

Having developed a reliable process for producing nano-dot array masks scalable to sub-10 nm dimensions, we next demonstrate their integration into advanced spin-photo-electronic devices, such as new types of nano-oscillators [9-14] and the newly proposed lasers [15]. The 1-10 nm scale is particularly interesting as it enables devices based on non-equilibrium injection, even for metals, due to the single-dot size being comparable or smaller than the characteristic relaxation length scales in the material. Fabricating and integrating sub-10 nm dot arrays into circuit-driven devices is a non-trivial task for any patterning technique (see Introduction) and, to our knowledge, has not been demonstrated to date.

For the spin-laser device of [15], for example, the bottom electrode must be thick to serve as an efficient electron and phonon bath under high-current injection. We take that into account in the design and start the structure with a sputter-deposited tri-layer of Al(180 nm)/$SiO_2$(15 nm)/Au(10 nm) onto a Si/$SiO_2$(500 nm) substrate. The bottom electrode, later to serve as one side of a 10-100 μm range optical resonator, is patterned using the standard optical lithography. Specifically, a double-layer resist LOR7B(500 nm)/S1813(1.5 μm) is spun onto the Au surface, thermally treated, mask-exposed, developed, ion milled for 2 h, and lifted-off in 1165 remover at 60ºC to form a 100 μm wide bottom electrode.

The sample is then cleaned in oxygen RIE for 2 min, which makes the surface hydrophilic, and covered with 2-3 drops of the colloid solution forming the polystyrene particle monolayer during the above spinning sequence. The particles are scaled down using the multi-step process detailed above to form a nano-hole array mask on the bottom electrode surface. RIE plasma etching with $CF_4$ is used for 2 min for making the contact through the 15 nm thick $SiO_2$ with the 180 nm thick bottom electrode of Al. The etching time for 15 nm of $SiO_2$ is 1 min. Using the chemical selectivity of the $CF_4$ etch to $SiO_2$, we etch twice longer (2 min) in order to form a good undercut in $SiO_2$, which is important for the following deposition steps.

Sputtering was used for deposition of the active point-contact region. The material combination was selected to represent the spin-injection laser device [15,24,25], and consisted of a spin-majority/minority ferromagnetic bi-layer $Fe_{0.7}Cr_{0.3}$(10nm)/Fe(15 nm) [26] capped with Cu(10 nm), for spin population inversion injection. We find that the angle of deposition through the 10-nm openings in the mask is an important parameter determining the size of the resulting point contacts – the closer to normal incidence the closer the resulting contact size to the mask feature size, and the sharper the angle of incidence the deeper sub-10 nm the nano-contacts are due to the double shadowing effect illustrated in Fig. 5 (direct mask-shadowing and shadowing from material buildup on the mask edge). We used two angles of deposition, near-normal incidence and approximately 45-degree incidence, and estimate that the average size of the nano-contacts obtained for the angled deposition was approximately 5 nm. Finally, surface protection for subsequent processing steps was done with two layers of Cr(5 nm)/Au(10 nm) deposited by e-beam evaporation. The key elements of the point-contact structure obtained are illustrated in the bottom panel of Fig. 5a.

The final integration step is photolithographic patterning of the top electrode, in the case of the demonstrator devices below in the shape of a photon resonator for IR-THz photons (Fig. 5a, top panel). The process is analogous to the one used for the bottom electrode, but employing a different photo-mask. The top electrode mask has different diameter disks and half-disks in the range of 10-50 microns. The pattern transfer is done by ion milling for 1 h. The etching time was calibrated using surface profilometry such as to stop the etch at the Al bottom electrode. The sample was then capped with a 40 nm thick $SiO_2$ layer for insulation, rotating the sample holder during deposition. Finally, the resist was lifted off, and the last step of lithography was using negative resist and depositing a 200 nm thick Al top electrode.

## Device examples

The focus of this letter is the new method of integrating sub-10 nm structures into nano-devices. We briefly demonstrate the method using two physical effects found in magnetic nano-contacts, namely, spin-magnon and spin-photon relaxation. The method is not limited to photonics or spintronics, however, and should have a wide application range in various types of physical systems.

We first estimate the expected circuit characteristics of our typical integrated point-contact array. For individual 5-10 nm metallic point contacts the resistance is essentially given by the geometry (the so-called Sharvin resistance [27]) and approximately is 20-10 Ω. For an ideal 10x10 μm$^2$ point-contact array with a 200 nm inter-contact spacing, the number of contacts is 2500. Therefore, the expected resistance of the array is of the order of 10 mΩ. A non-ideal array would have fewer contacts and therefore higher resistance. Over-etching the polystyrene monolayer and sharp-angle deposition of the core material, as discussed in detail above, can result in only a fraction of the array actually connected and the individual nano-contact size substantially smaller than the mask feature size of 10 nm, as measured by SEM and AFM. In this limit we are able to reach the array resistance range of the order of 1 Ω.

We have prepared test samples with the point-contact core made of a single nonmagnetic metallic element (Cu with a Cr underlayer), where no effects due to spin-flip relaxation are expected, only phonon relaxation (heat). The typical array resistance is measured to be 10-20 mΩ. The current-voltage characteristic is smooth and approximately parabolic, typical of the expected phonon background. Thus, these test data agreed with the expected behavior and showed that the fabricated nano-contact arrays are of high quality.

We next demonstrate a point-contact array with the contacts having a magnetic core. More specifically, the core material is a majority/minority ferromagnetic bi-layer of Fe/Fe$_{0.7}$Cr$_{0.3}$ [26], where due to the opposite spin-polarizations of the two materials at the interface a strong spin accumulation is expected. The top contact is a half-disk-shaped 10 μm radius Al electrode. Figure 5b is a resistance versus bias current characteristic for the device and shows a clear current-induced hysteretic switching, typical of magnetic point contacts [28,29], superposed on the phonon background. The mechanisms behind is formation of atomic/nano-scale domain walls in the nano-constriction under the spin-transfer torque (STT) from the spin accumulation at the Fe/Fe-Cr interface. The switching in both directions occurs at one bias polarity, which is characteristic of the STT effect. The change in resistance is approximately 2%, typical of domain wall magneto-resistance. The array resistance is of the order of 10 mΩ, consistent with the expected range for a nearly fully connected sub-10 nm point-contact array. Thus, we demonstrate the STT effect in the fabricated nano-devices. The extremely regular array layout with the extremely small contact size, as well as the relative ease of the colloidal monolayer process, should make this structure very promising for GHz nano-oscillators [9-14]. Optimization would involve substituting improved spin-valve materials for the core region and, if needed, tuning the array lattice spacing to achieve better spin-wave modes interference.

Another interesting application of the spin-majority/minority $Fe/Fe_{0.7}Cr_{0.3}$ contact-core material used above is the spin-flip photon emission effect [15,24,25], which requires spin-polarized and non-equilibrium in energy injection and, therefore, near-ballistic point-contact arrays. To achieve this we have performed fabrication at the limit in the etching and angle-deposition parameters discussed above, for deep sub-10 nm point-contact size and a smaller operating fraction of the array, so that the injection voltage per contact is greater than the exchange splitting in the ferromagnetic point contact core (10-100 mV, see [24,25] more details). Such high-bias, high-density, spin-polarized injection produces large non-equilibrium spin accumulation in the contact core, which allows spin-flip photon-emission transitions, vertical in electron's momentum. A photon emitted by a spin-flip process is contained within the resonator and enhanced by the high-dielectric constant, high-transparency $SiO_2$ oxide matrix [15]. The lifetime of the emitted photon is long due to the high transparency of the oxide, so the photon has a high probability to stimulate another spin-flip transition. At a critical bias, a cascading avalanche of stimulated spin-flip photon-emitting transitions - a laser action - takes place, generating high-density electromagnetic modes in the resonator. This critical photon-emission threshold must manifest itself as threshold-type changes in the current-voltage characteristics of our fully enclosed optical resonator. Such threshold-type excitations, of giant magnitude, are indeed observed in the device resistance, as shown in Fig. 5c. Tis demonstration opens the way to explore a new type of THz lasers based on spin-flip photon emission.

## Conclusions

Colloidal patterning in the form of large-area hexagonal-lattice arrays is demonstrated to scale down to sub-10-nm dimensions in the feature size. This is comparable or smaller than the key relaxation lengths in various materials including metals, which enables a wide range of new applications in nanotechnology. Large-area, near-ballistic-injection point-contact arrays are used to demonstrate integration of the developed nanofabrication technique into new types of spintronic and photonic devices.

**Figure captions**

1. Tapping mode atomic force microscopy images of a typical monolayer, with the particle diameter (and the inter-particle distance) of 200 nm on a large scale (a) and small scale (b). Scanning electron microscopy image of the sample (c) shows that the nanoparticle array has a nearly perfect close-packed hexagonal lattice.

2. Illustration of the four steps in down scaling the particles of a polystyrene monolayer to the 10-nm range, which later is to serve as a nano-array mask. (a) Forming a self-assembled hexagonal-close-packed monolayer. (b) Reactive oxygen plasma with ICP is used to reduce the size of the particles. (c) A reinforcing layer of Aluminum is deposited to serve as a hard mask. (d) A lift-off of the particles by etching completes the fabrication of the nano-mask.

3. Polystyrene nanoparticle monolayer after plasma-ICP etching for 165 (a), 180 (b), and 195 seconds (c). (d) AFM height profiles of the three samples shown in (a,b,c) – raw data, without de-convolution.

4. SEM images of a typical colloidal-monolayer mask, shown on four scales in (a-d). The average inter-dot distance is 200 nm (c,d), corresponding to the original polystyrene particle diameter. The average dot size is variable by the plasma-ICP etching time, and reaches down the sub-10 nm (c). (d) Shows a mask close-up with a nearly perfect hexagonal -close-pack pattern, with the lattice constant of 200 nm and the hole diameter of 13 nm.

5. (a) Illustration of the integrated circuit device using a colloidal nano-array mask and two steps of lithography for the bottom and top electrodes. Top panel shows the layout of the device, with the top electrode formed in the shape of a photon resonator for IR-THz, 10-50 μm diameter. Bottom panel shows the nano-array mask and the core structure of a single sub-10 nm point contact. (b) Resistance versus bias current characteristic for a Fe/Fe$_{0.7}$Cr$_{0.3}$ point-contact array (low resistance range, 10-20 mΩ), showing current-induced hysteretic switching superposed on to parabolic phonon background. (c) Resistance versus bias current for a 36 μm-diameter spin-flip laser array (high-resistance limit, ~1 Ω), showing giant excitations of threshold type at critical pumping consistent with the onset of the expected of stimulated emission in the device [15].

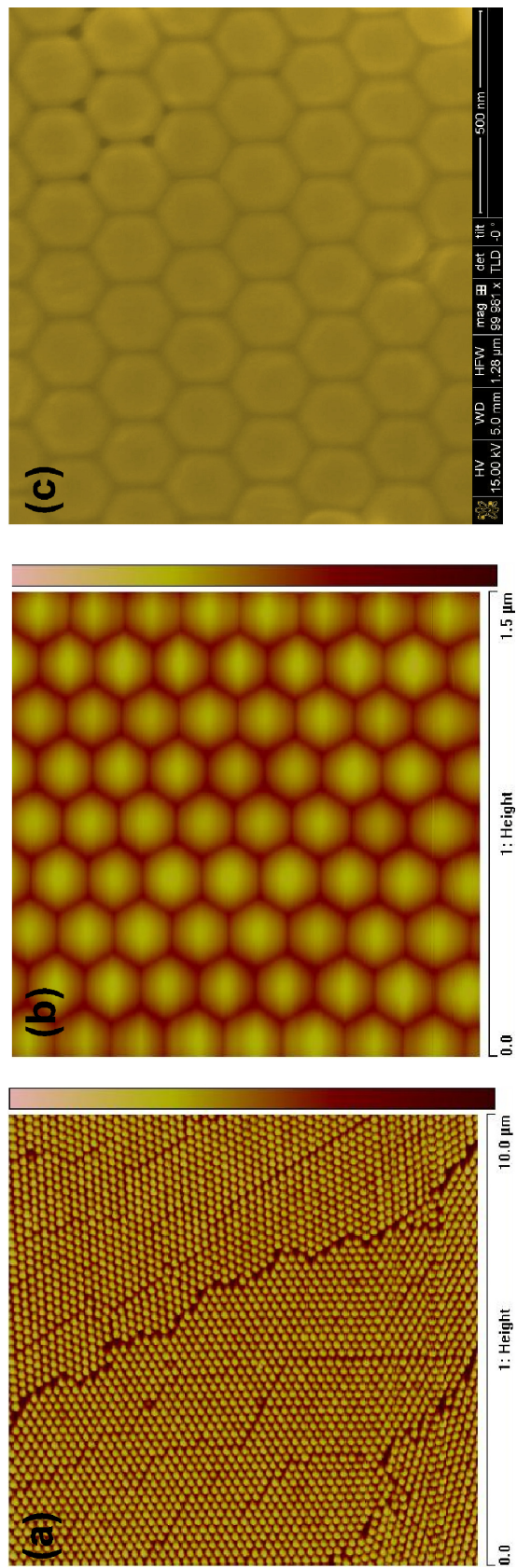

**Fig. 1**

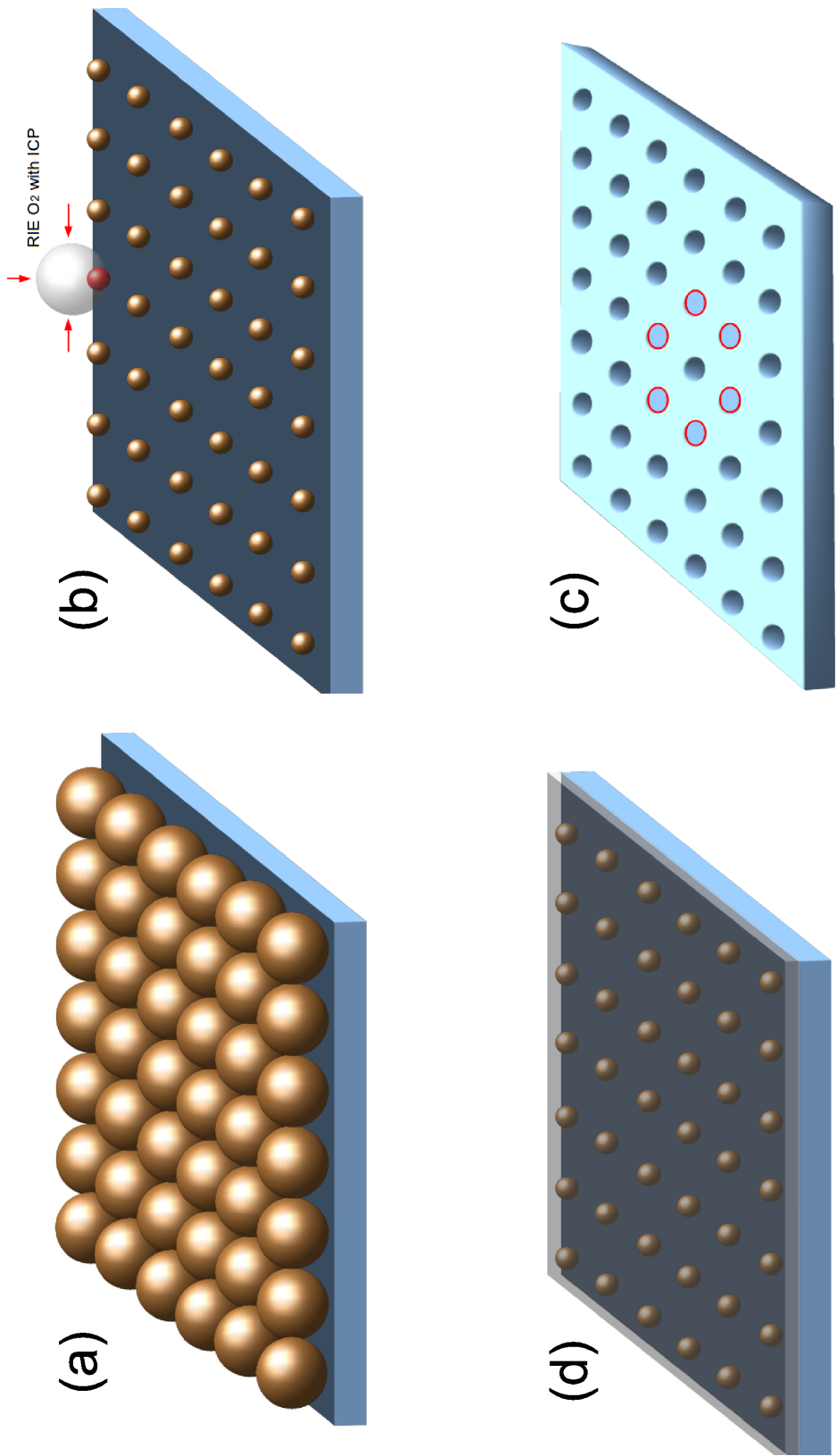

**Fig. 2**

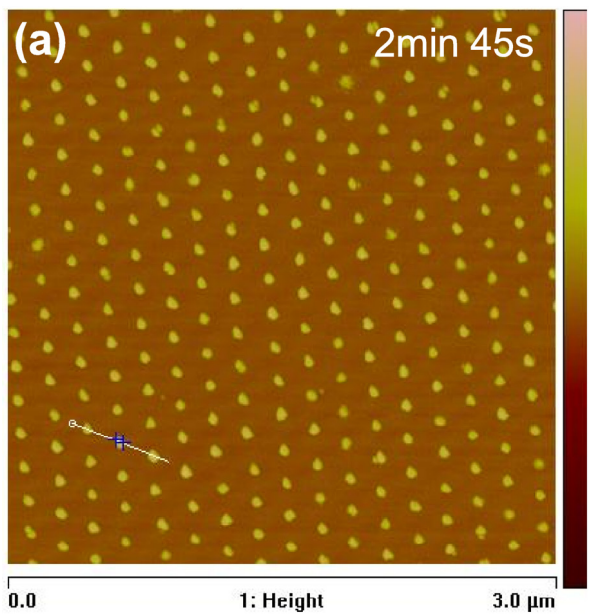
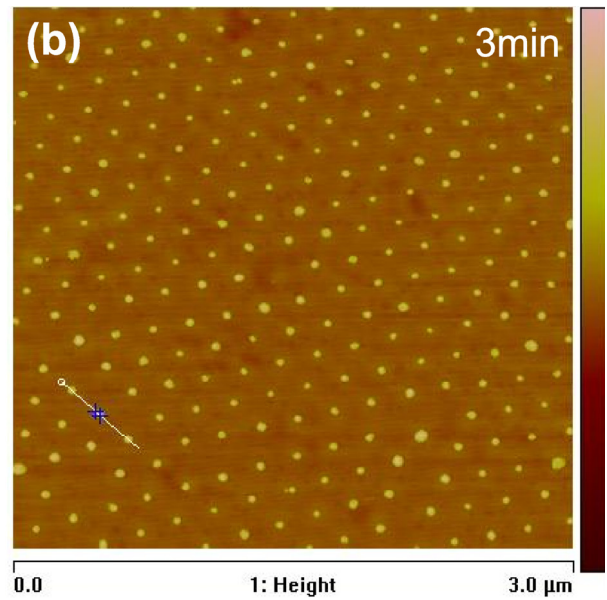
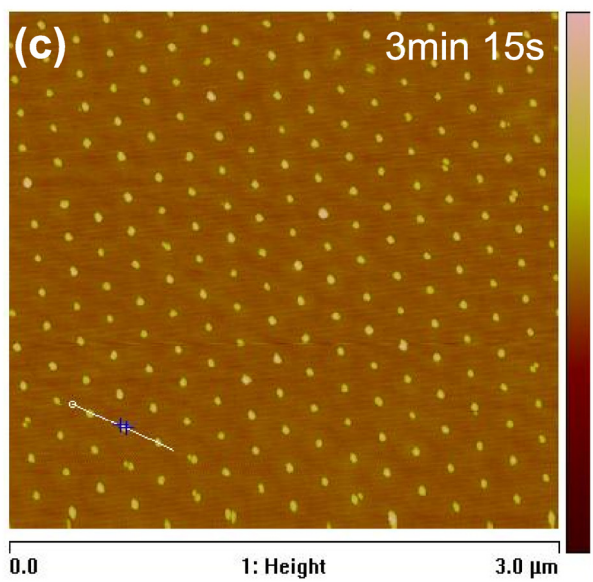
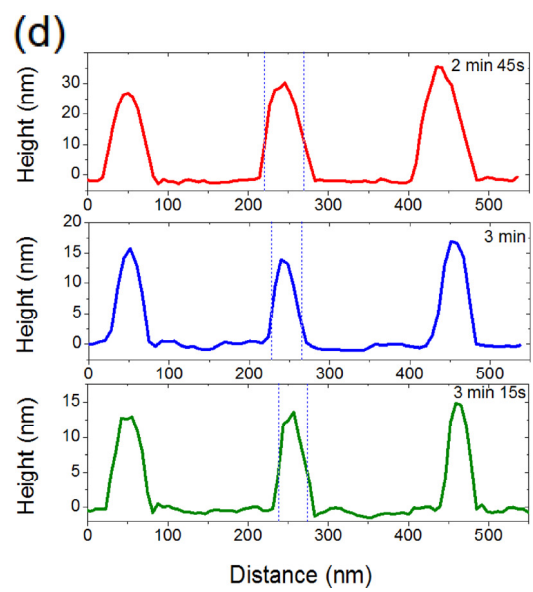

**Fig. 3**

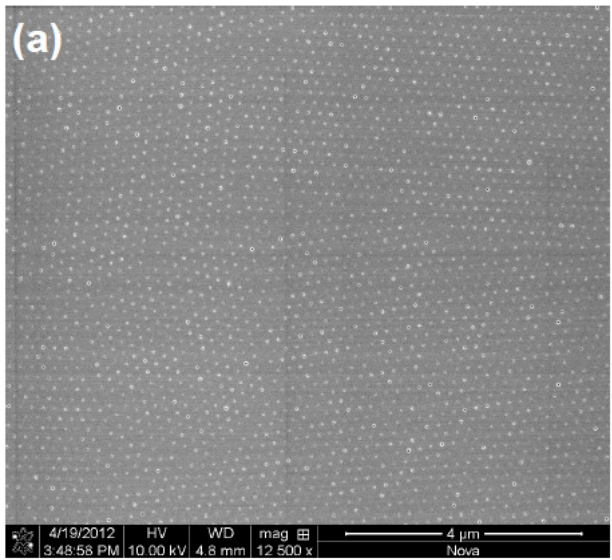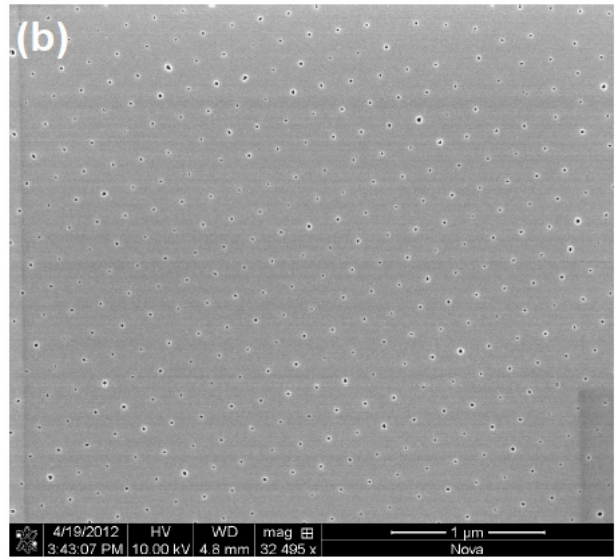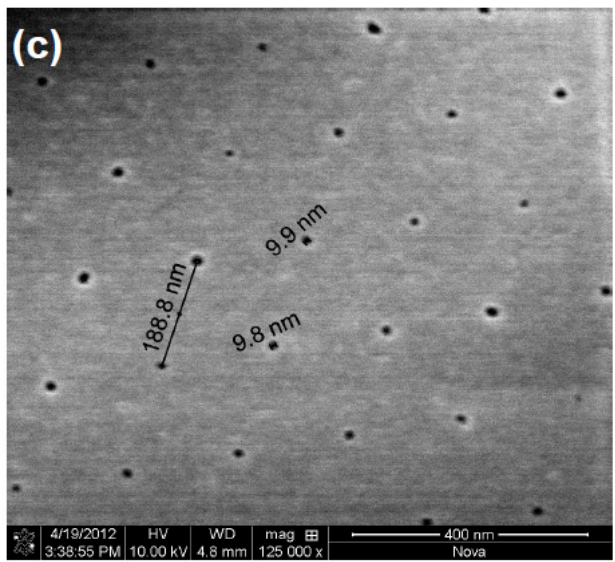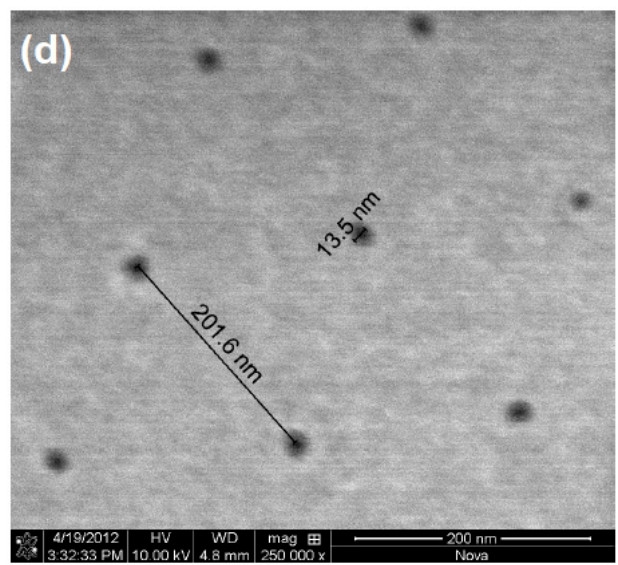

**Fig. 4**

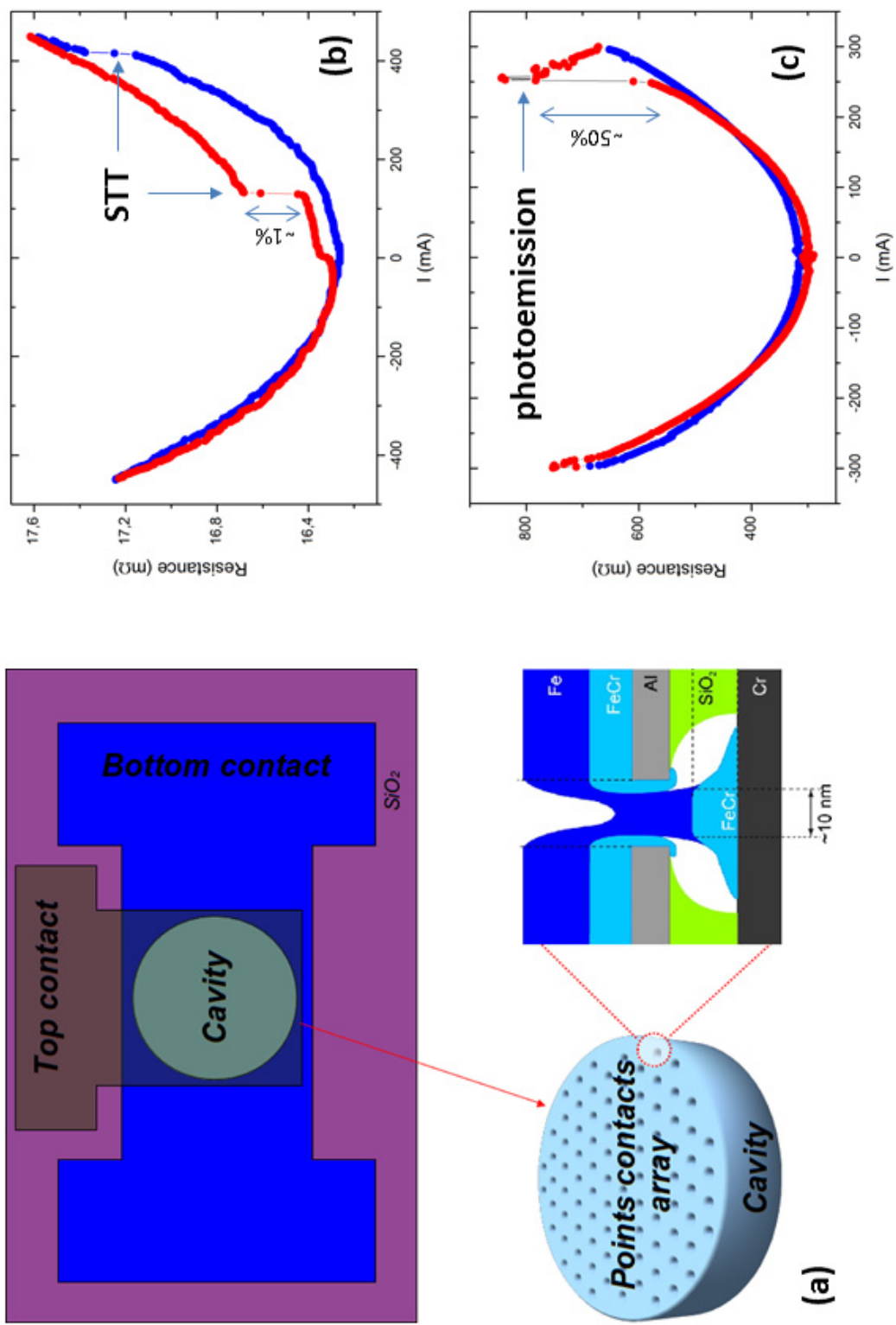

Fig. 5